\documentclass[prb,twocolumn,superscriptaddress,msmath,amssymb]{revtex4}

\usepackage{psfig,bm}
\begin{document}
\psfigurepath{./figures}

\title{Slow spin-glass and fast spin-liquid components in quasi-two-dimensional La$_{2}$(Cu,{\it Li})O$_4$}

\author{Y. Chen}
\affiliation{Condensed Matter and Thermal Physics, Los Alamos National Laboratory, Los Alamos, NM 87545}
\affiliation{NIST Center for Neutron Research, National Institute of Standards 
and Technology, Gaithersburg, MD 20899}
\affiliation{Dept.\ of Materials Science and Engineering, University of
Maryland, College Park, MD 20742}
\author{Wei Bao}
\email{wbao@lanl.gov}
\affiliation{Condensed Matter and Thermal Physics, Los Alamos National Laboratory, Los Alamos, NM 87545}
\author{Y. Qiu}
\affiliation{NIST Center for Neutron Research, National Institute of Standards 
and Technology, Gaithersburg, MD 20899}
\affiliation{Dept.\ of Materials Science and Engineering, University of
Maryland, College Park, MD 20742}
\author{J. E. Lorenzo}
\affiliation{CNRS, BP 166X, F-38043, Grenoble, France}
\author{J.L. Sarrao}
\affiliation{Condensed Matter and Thermal Physics, Los Alamos National Laboratory, Los Alamos, NM 87545}
\author{D. L. Ho}
\affiliation{NIST Center for Neutron Research, National Institute of Standards 
and Technology, Gaithersburg, MD 20899}
\affiliation{Dept.\ of Materials Science and Engineering, University of
Maryland, College Park, MD 20742}
\author{Min Y. Lin}
\affiliation{NIST Center for Neutron Research, National Institute of Standards 
and Technology, Gaithersburg, MD 20899}
\affiliation{ExxonMobil Research and Engineering Company, Annandale, NJ 08801}

\date{\today}

\begin{abstract}
In conventional spin glasses, magnetic interaction is not strongly anisotropic 
and the entire spin system is believed to be frozen below the spin-glass 
transition temperature. In La$_2$Cu$_{0.94}$Li$_{0.06}$O$_4$,
for which the in-plane exchange interaction dominates the interplane one, 
only a fraction of spins with antiferromagnetic correlations extending to 
neighboring planes become spin-glass. The remaining spins with only in-plane 
antiferromagnetic correlations remain spin-liquid at low temperature. 
Such a novel partial spin freezing out of a two-dimensional
spin-liquid observed in this cold neutron scattering study 
is likely due to a delicate balance between disorder and quantum fluctuations
in the quasi-two dimensional $S$=1/2 Heisenberg system.
\end{abstract}


\maketitle

\section{INTRODUCTION}

The parent compound for high transition-temperature superconductors,
La$_2$CuO$_4$, is an antiferromagnetic insulator. Magnetic exchange
interaction $J$ between the nearest neighbor $S$=1/2 spins of Cu$^{2+}$ ions
in the CuO$_2$ plane is several orders of magnitude stronger than the
interplane exchange interaction, making quantum spin fluctuations an
essential ingredient for magnetic properties in the quasi-two-dimensional 
(2D) Heisenberg system\cite{la2dv2,2dheis,2dheiqc}. 
The N\'{e}el temperature $T_N$ of La$_2$CuO$_4$ is suppressed rapidly 
to zero by $x_c=2$$-$3\% hole dopants such as Sr, Ba or Li\cite{ht_pd,nagano,Li214phs}, 
while it is suppressed with isovalent Zn
substitution at a much higher concentration close to the
site dilution percolating threshold of $\sim$30\%\cite{ZnSr214}. 
The strong effect of holes has been shown to be related to induced
magnetic vortices, which are topological defects in
2D systems\cite{h_haas,h_ctb}. The paramagnetic phase exposed by hole doping
at $T \ll J/k_B$ is dominated by zero-point quantum spin 
fluctuations and is referred to as a quantum spin liquid\cite{2dheis}.
Detailed predictions for spin dynamics have been made for the quantum 
spin-liquid\cite{2dheis,2dheiqc}.

However, in a wide doping range of La$_{2}$Cu$_{1-x}$Li$_x$O$_4$
below $\sim$10~K, a spin-glass transition has 
been reported in muon spin rotation ($\mu$SR)\cite{Li214phs}, 
nuclear quadrupole resonance (NQR)\cite{Li214NQR} and 
magnetization\cite{Li214phs2} studies.
A similar magnetic phase diagram has also been reported for
La$_{2-x}$Sr$_x$CuO$_4$ and 
Y$_{1-x}$Ca$_x$Ba$_2$Cu$_3$O$_6$\cite{sg_drh,bjbquasi,fchou,muchn,gqcp,sg_fcc}.
In conventional spin glasses, magnetic interactions are more or
less isotropic in space, and the entire spin system is believed 
to be frozen in the spin-glass phase\cite{byoung}. 
Such was also the conclusion of a comprehensive magnetization study 
on La$_{2-x}$Sr$_x$CuO$_4$\cite{sg_fcc}. Although magnetization
can only account for a tiny fraction of spins, theoretical
pictures were proposed for spin-freezing in the whole 
sample\cite{sg_fcc,sg_rjg}.
If the spin-glass phase in hole-doped cuprates behaved as in 
conventional spin-glasses, the ground state would be a spin-glass, instead of 
the N\'{e}el order for doping smaller than $x_c$, or a quantum spin liquid
for doping larger than $x_c$. Thus, as pointed out
by Hasselmann et~al.\cite{sg_cn}, the quantum critical point of
the antiferromagnetic phase at $x_c\approx 2$$-$3\% would be preempted.

In widely circulating ``generic'' phase-diagram for laminar cuprates,
the ``reentrant'' spin-glass transition below the N\'{e}el temperature
is generally ignored. Also generally ignored is the spin-glass transition below the superconducting transition. The spin glass
phase exists side by side with the N\'{e}el order at lower doping and the superconducting order at higher doping in this neat picture. 
This ``generic'' picture does not conform to experimental results,
and serves to support the theory that the spin-freezing is an extrinsic dirt effect. However, there are other theories which consider spin-freezing
intrinsic to the doped cuprates\cite{slf,sg_sahc}.
Physical quantities in the doping regime, including spin excitation
spectra, have also been calculated from microscopic 
model\cite{march1,march2}.

Recently, 2D spin fluctuations in La$_{2}$Cu$_{1-x}$Li$_x$O$_4$ 
($0.04\leq x\leq 0.1$) were observed to remain liquid-like below  
the spin-glass transition temperature\cite{bao02c,bao04a}, $T_g\sim 9$~K,
which can be reliably detected using the $\mu$SR technique\cite{Li214phs}. 
The characteristic energy of 2D spin fluctuations 
saturates at a finite value below $\sim$50~K\cite{bao02c,bao04a}, as expected for a 
quantum spin liquid\cite{2dheis}, instead of becoming zero 
at $T_g$ as for spin-glass materials\cite{byoung,FeAl_msm}.
To reconcile these apparently contradicting experimental results, 
we have conducted a thorough magnetic neutron scattering investigation 
of La$_2$Cu$_{0.94}$Li$_{0.06}$O$_4$ to search for spin-glass
behavior. We found that in addition to the 
liquid-like 2D dynamic spin correlations, the rest of spins which participate 
in almost 3D and quasi-3D correlations become frozen in the 
spin-glass transition. This partial spin freezing in the laminar
cuprate is distinctly different from total spin freezing
in conventional 3D spin-glass materials.
The observed phase separation into spin glass and spin liquid components 
of {\em different dimensionality} sheds light on a long-standing confusion 
surrounding the magnetic ground state in hole-doped cuprates.

The remaining of the paper is organized as the follows.
Section II covers experimental details concerning the sample and neutron scattering instrumentation. Section III covers small angle neutron
scattering, which is the ideal tool to detect ferromagnetic spin
clusters proposed in some theories for the spin-freezing state.
Section IV covers cold neutron triple-axis measurements. The
excellent energy resolution is important for this study. Finally,
in Section V, we discuss and summarize our results.

\section{EXPERIMENTAL DETAILS}

The same single crystal sample of La$_2$Cu$_{0.94}$Li$_{0.06}$O$_4$
used in the previous higher energy study\cite{bao02c} was investigated in this work. 
$T_g$$\approx$8~K was determined in $\mu$SR study\cite{Li214phs}
and is consistent with magnetization work\cite{Li214phs2}. 
The lattice parameters of the orthorhombic $Cmca$ unit cell are
$a$=5.332$\AA$, $b$=13.12$\AA$ and $c$=5.402$\AA$ at 15~K. 

Wave-vector
transfers {\bf q} near (000) and (100) in both the ($h0l$) and ($hk0$) 
reciprocal planes were investigated at NIST using the 30 meter high 
resolution small angle neutron scattering (SANS) instrument at NG7, 
and cold neutron triple-axis spectrometer SPINS.
We set the array detector of NG7-SANS to  
1 and 9 m, corresponding to a $q$ range from 0.012 to 0.39~$\AA^{-1}$ 
and from 0.0033 to 0.050~$\AA^{-1}$, respectively.  
At SPINS, the (002) reflection of pyrolytic 
graphite was used for both the monochromator and analyzer. Horizontal 
Soller slits of 80$^{\prime}$ were placed before and after the sample.
A cold BeO or Be filter was put before the analyzer to eliminate 
higher order neutron in the fixed $E_f$=3.7 or 5 meV configuration,
respectively.

Sample temperature was controlled by a pumped $^4$He cryostat which could
reach down to 1.5~K.

\section{Small angle neutron scattering}

Hole induced ferromagnetic exchange has been theoretically proposed
in the CuO$_2$ plane\cite{sg_rjg,sg_cn}. It is regarded as competing with
the original antiferromagnetic exchange, thus, leading to the spin-glass
transition.
Although long-range ferromagnetic order has never been observed,
there is the possibility of short-range ferromagnetic spin clusters 
which freeze in the spin-glass state in this class of spin-glass 
models\cite{sg_rjg,sg_lin}. SANS has been demonstrated as a powerful tool to probe such clusters\cite{FeAl_msm}.

Two reciprocal zones of La$_2$Cu$_{0.94}$Li$_{0.06}$O$_4$
were studied, with incident 
beam parallel to the (001) or (010) direction. Therefore, 
any spin orientation in the sample can be detected in our experiment. The
experiments were carried out at 3, 10, 15, 30 and 80~K.  A
collection time of 1 or 2 hours per temperature provides good
statistics. 

\begin{figure}[t]
\vskip -1ex
 \centerline{
\psfig{file=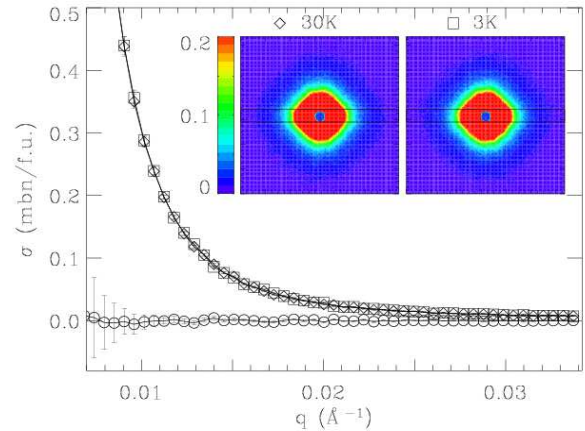,width=\columnwidth,angle=90,clip=}}
\vskip -3ex 
\caption{ \label{fig1} (color)
Measured SANS cross-section in the 10 pixels
wide rectangle shown in the inset at 3~K (squares) and 30~K
(diamonds), and the difference of the intensity between 3 K and 30 K
(circles) as a function of wave vector transfer $q$. Lines
are guide to the eye. Inset: the SANS pattern at 30 and 3~K,
with the intensity color scale at the left.  
}
\end{figure}

No temperature dependence in the scattering patterns 
could be detected. The inset to Fig.~\ref{fig1} shows SANS patterns at
3 K and 30 K with incident beam parallel to the (001) direction.
Intensity at 3 and 30~K in the rectangular box on the 
SANS pattern is shown in the main frame.
The difference intensity (circles) fluctuates around zero,
and its standard deviation sets a upper limit of 
1.5$\times$10$^{-7}$ bn or
1.4$\times$10$^{-3} \mu_B$ per Cu for ferromagnetic moments 
in the clusters.  

This result provides serious constrain on the class of theoretical
models for the spin-glass transition in doped cuprates\cite{sg_rjg}
which lead to formation of ferromagnetic clusters. Instead of
this ``large spin fixed point'', models leading to other fixed points
such as ``Griffiths fixed point'' as discussed by Lin et al.\cite{sg_lin} may be considered.

\section{triple-axis neutron scattering}

While no appreciable ferromagnetic signal was detected for
La$_2$Cu$_{0.94}$Li$_{0.06}$O$_4$, as in other Li-doped 
La$_2$CuO$_4$\cite{bao99a,bao04a}, antiferromagnetic
scattering was readily observed along the rods 
perpendicular to the CuO$_2$ plane and intercepting the 
plane at the commensurate ($\pi,\pi$)-type Bragg points 
of the square lattice. This means that antiferromagnetic
correlations in the CuO$_2$ plane are chessboard-like,
which is similar to electron-doped La$_2$CuO$_4$\cite{edoped1,edoped2},
but different from the more complex, incommensurate ones 
in La$_{2-x}$Sr$_x$CuO$_4$ at similar hole doping\cite{Waki_Sr3}.

\begin{figure}[t]
\vskip -1ex 
\centerline{
\psfig{file=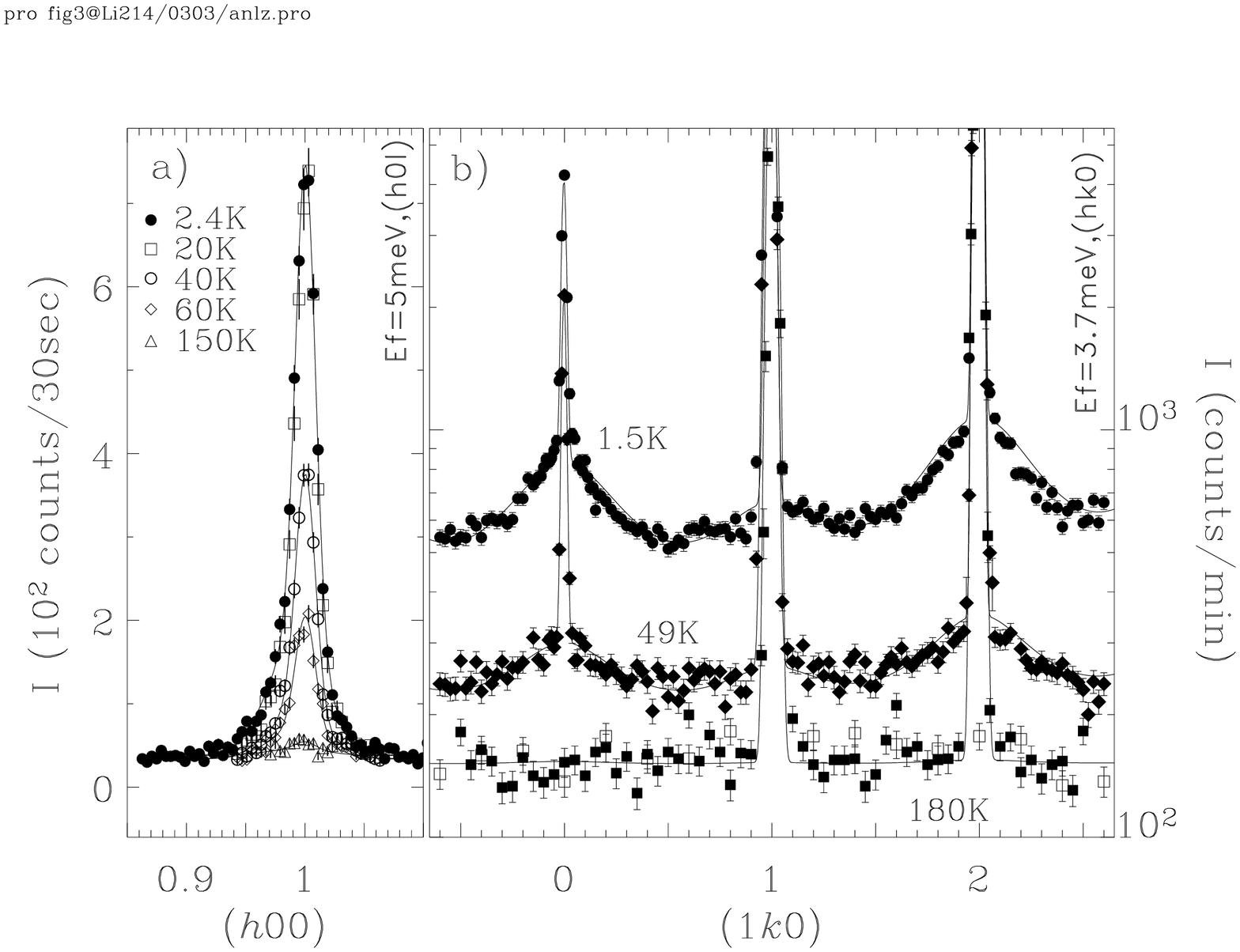,width=\columnwidth,angle=90,clip=}}
\vskip -2ex 
\caption{ \label{fig2} 
Representative magnetic quasielastic and elastic scattering along a) an
in-plane direction and b) the interlayer direction between 1.5 and 180~K.
Open squares in b) were measured at (1.06,$k$,0) 
and represent background.
The solid lines are resolution convoluted 
$S^{3D}({\bf q},E)+S^{q3D}({\bf q},E)$ in Eq.~(\ref{eq1})-(\ref{eq2}).
}
\end{figure}

Scans through such a rod in the CuO$_2$ plane at various temperatures 
with the SPINS spectrometer set at $E$=0 are shown in Fig.~\ref{fig2}(a). 
Inelastic scans have been reported previously in a related but different study which focuses on scaling in different quantum regimes\cite{bao02c}. There is little change in the peak width
in these scans, consistent with previous results of 
temperature independent in-plane correlation length 
for La$_2$Cu$_{0.95}$Li$_{0.05}$O$_4$\cite{la2dv2}
and La$_{2-x}$Sr$_x$CuO$_4$ (0.02$\leq x\leq$0.04)\cite{la2bkb}
below 300~K. Modeling the width of the rod in 
Fig.~\ref{fig2}(a) with Lorentzian
\begin{equation}
\mathcal{L}^{\xi}(q)=\frac{\xi}{\pi[1+(q\xi)^2]},
 \label{eq0}
\end{equation}
the {\em lower limits} from deconvolution is 
$\xi_{\Box}\ge 274 \AA$, where the  
$\Box$ indicates the correlation length as in-plane.
These large antiferromagnetic clusters in the CuO$_2$ plane correlate in three different 
ways in the interlayer direction, giving rise to
almost 3D, quasi-3D and 2D magnetic correlations. 
Let us now examine the three components.

Scans along the rod in the interlayer 
direction, with the SPINS spectrometer set at $E$=0, 
were measured at various temperatures
from 1.5 to 180~K. A few of them, at 1.5, 49 and 180~K, respectively,
are shown in Fig.~\ref{fig2}(b). 
Magnetic intensity is composed of both broad and sharp
peaks at magnetic Bragg points (100) and 
(120) of the parent compound. 
The (110) peak is temperature-independent thus nonmagnetic.
Fitting the broad
peaks to Eq.~(\ref{eq0}), we obtained an interlayer correlation 
length $\xi^{q3D}= 6.2(4)$~$\AA$. Again, no temperature dependence
can be detected for $\xi^{q3D}$ below 49~K. Above 49~K, signal is
too weak to have a reliable determination of $\xi^{q3D}$.
Thus, the quasi-3D spin correlations are typically three planes thick. 
For the sharp peak at (100) or (120), only the {\em lower limit} 
for the correlation length can be reliably estimated: 
$\xi^{3D}\ge 168\AA$, since the width is close to instrumental resolution. 
Therefore, the number of correlated antiferromagnetic planes is more 
than 50, resembling a 3D antiferromagnetic order.

\begin{figure}[t]
\vskip -5ex
\centerline{
\psfig{file=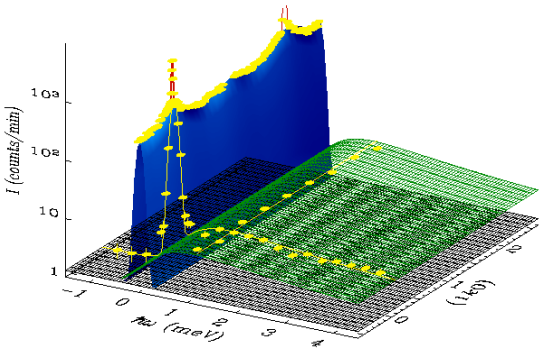,width=1.1\columnwidth,angle=90,clip=}} 
\vskip -4ex
\caption{ \label{fig3} (color)
Measured $S({\bf q},E)$ as a function of $E$ and interlayer $k$ at 1.5~K
with a logarithmic intensity scale, 
showing three color-coded magnetic components in Eq.~(5). 
$S^{3D}({\bf q},E)$ (red) and $S^{q3D}({\bf q},E)$ (blue) 
are energy-resolution limited
at $E=0$, representing very slow spin dynamics which is associated with
the spin-glass freezing. They are modulated along the interlayer $k$
direction. The spin-liquid component, $S^{2D}({\bf q},E)$ (green), 
has a finite energy scale of about 1~meV
below 50~K, and 0.18$k_B T$ above 50~K\cite{bao02c}. 
It is flat along the $k$ direction.
A few representative scans at 1.5~K are shown with yellow symbols. The black
surface indicates background of $\sim$1.3 counts/min.
}
\end{figure}

Both the broad and sharp peaks in Fig.~\ref{fig2}(b)
are energy-resolution-limited
with the half-width-at-half-maximum $=$~0.07 meV. 
The energy scan in Fig.~\ref{fig3} is an example and more
can be found in Fig.~2 in reference [\cite{bao02c}]. 
However, these peaks should not be regarded automatically
as from {\em static} magnetic order. Static magnetic signal 
was observed only below $T_g$=8~K at the spin glass transition 
in $\mu$SR study\cite{Li214phs}, which has a much better energy
resolution. Thus, the quasi-3D and almost 3D  
correlations are slowly dynamic for $T> 8$~K, with their spectra
faster than 1 MHz\cite{Li214phs,bjbquasi}, the zero-field $\mu$SR static cutoff frequency, but slower than 17 GHz=0.07 meV/$h$, the frequency 
resolution at spectrometer SPINS.

The 2D antiferromagnetic correlations have been investigated 
in detail\cite{bao02c}. The dynamic magnetic structure factor,
\begin{equation}
S^{2D}({\bf q},E)=\sum_{\bm{\tau}} 
\mathcal{L}^{\xi_{\Box}}(\bm{\kappa}_{\Box})
\frac{\chi''(E)}{\pi \left(1-e^{-\hbar\omega/k_BT}\right)},
\label{eq4}
\end{equation}
where ${\bm{\tau}}$ is a magnetic Bragg wave-vector
and $\bm{\kappa}\equiv {\bf q}-{\bm{\tau}}$, has been determined from measurements in the energy range, $E \le 4.2$~meV, between 1.5 and 
150~K.
Eq.~(\ref{eq4}) is independent of $k$, befitting to a 2D magnetic
correlation, see the flat $k$ scan at 1.2 meV in Fig.~\ref{fig3}.
The almost 3D and quasi-3D spin correlations described in previous
paragraphs can be written as
\begin{equation}
S^{3D}({\bf q},E)=I^{3D} \sum_{\bm{\tau}} 
\mathcal{L}^{\xi_{\Box}}(\bm{\kappa}_{\Box})
\mathcal{L}^{\xi^{3D}}(k-\tau_k) \mathcal{L}^{1/\epsilon}(E) \label{eq1}
\end{equation}
and
\begin{equation}
S^{q3D}({\bf q},E)=I^{q3D} \sum_{\bm{\tau}} 
\mathcal{L}^{\xi_{\Box}}(\bm{\kappa}_{\Box}) 
\mathcal{L}^{\xi^{q3D}}(k-\tau_k) \mathcal{L}^{1/\epsilon}(E), \label{eq2}
\end{equation}
respectively, where $\epsilon < 0.07$ meV, the spectrometer energy resolution.
Note that we use conventional Lorentzian function, Eq.~(\ref{eq0}),
to model sharp peaks which we could not experimentally resolve,
in addition to $\mathcal{L}^{\xi^{q3D}}$ in Eq.~(\ref{eq2}) which we could
resolve. We are fully aware that the true
peak profile can be different for these unresolved peaks. 
The use of Eq.~(\ref{eq0})
is for the purpose of calculating resolution convolution of 
Eq.~(\ref{eq4})-(\ref{eq2}), which is used in the following paragraphs to 
obtain correct normalization of $I^{2D}$, $I^{q3D}$ and $I^{3D}$. 
The choice of the function will
not affect the result as long as the function describes a sharp
peak significantly narrower than instrument resolution.

With negligible ferromagnetic correlations, the total
dynamic structure factor is a summation of Eq.~(\ref{eq4})-(\ref{eq2}),
\begin{equation}
S({\bf q},E)=S^{2D}({\bf q},E)+S^{q3D}({\bf q},E)+S^{3D}({\bf q},E). \label{eq5}
\end{equation}
Of the four variables of $S({\bf q},E)$, {\bf q}$_{\Box}$ are fixed
at the ($\pi,\pi$)-type Bragg points by the sharply peaked 
$\mathcal{L}^{\xi_{\Box}}(\bm{\kappa}_{\Box})$\cite{note}.
To comprehend the composition of $S({\bf q},E)$, it is 
sufficient to plot $S({\bf q},E)$ as a function of $E$ and the 
interlayer wavenumber $k$. Such a plot of measured $S({\bf q},E)$
at 1.5~K is shown with a logarithmic intensity scale in Fig.~\ref{fig3}.
The temperature and {\bf q} independent incoherent scattering 
and other background at $E=0$ has been subtracted, which can be 
determined, e.g., by the 180~K scan in Fig.~\ref{fig2}(b).
The sharp peak fitted by the red curve is from $S^{3D}({\bf q},E)$,
the narrow blue ridge at $E$=0 from $S^{q3D}({\bf q},E)$, and
the green surface from $S^{2D}({\bf q},E)$. The red peak at (100)
is about one order of magnitude stronger than the peak intensity
of the blue surface, and three orders of magnitude stronger 
than the peak intensity of the green surface. Thus, $S^{3D}({\bf q},E)$
is the easiest component to be observed in a neutron scattering experiment,
and is often mistakenly attributed to a {\em static} magnetic order.

\begin{figure}[t]
\centerline{
\psfig{file=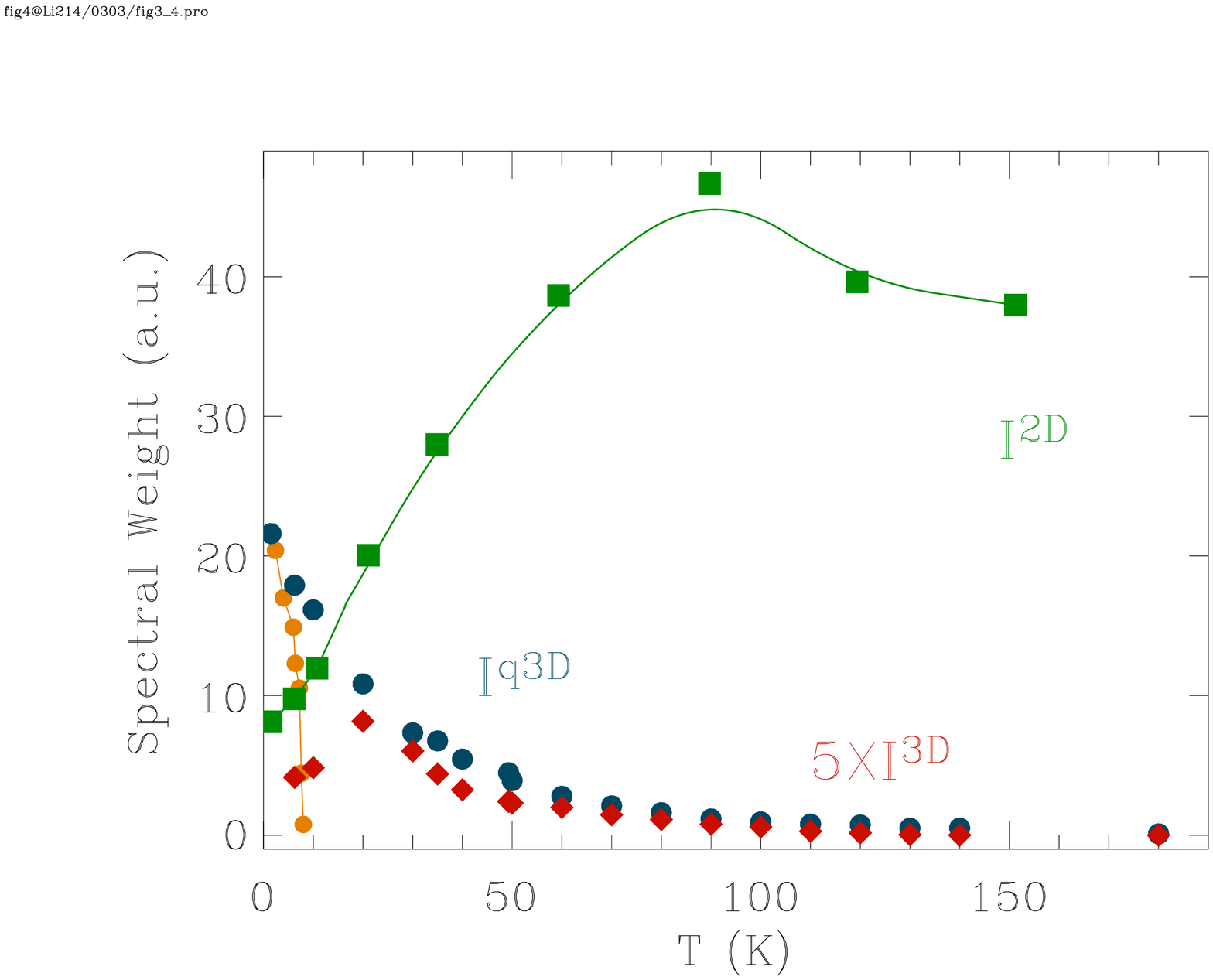,width=\columnwidth,angle=90,clip=}} 
\vskip -2ex
\caption{ \label{fig4} (color)
Temperature dependence of spectral weights
$I^{2D}$ (green), $I^{q3D}$ (blue) and $I^{3D}$ (red) in the same unit for three experimentally separable antiferromagnetic components 
in Eq.~(\ref{eq5}). $I^{3D}+I^{q3D}$ is the total spectral weight of 
the spin-glass component.
$I^{2D}$ is the spectral weight of the spin-liquid component 
within $|E|< 10$ meV, thus the lower limit of its total spectral weight. The orange circles represent squared {\em static} order parameter 
of the spin-glass transition, which was measured by 
$\mu$SR\cite{Li214phs} and equals to $I^{3D}+I^{q3D}$
at T=0.}
\end{figure}

The spectral weights 
$\int d{\bf q} dE\, S^{3D}({\bf q},E)$$\equiv$$I^{3D}$ and 
$\int d{\bf q} dE\, S^{q3D}({\bf q},E)$$\equiv$$I^{q3D}$ can be obtained
by fitting resolution-convoluted Eq.~(\ref{eq1})-(\ref{eq2})
to scans such as those shown in Fig.~\ref{fig2}(b). 
They are shown as a function of 
temperature in Fig.~\ref{fig4}, with $I^{3D}$ magnified by a 
factor of 5 for clarity. For the 2D component, the spectral weight is
\begin{equation}
I^{2D}\equiv \int dE\, \frac{2\chi''(E)}{\pi \left(1-e^{-\hbar\omega/k_BT}\right)},
 \label{eq6}
\end{equation}
where the integration limits are $\pm \infty$.
Green squares in Fig.~\ref{fig4} represent the lower bound of
$I^{2D}$ with the energy integration limited in 
$|E| \le 10$~meV, using the analytical expression 
of $\chi''(E)$ in reference [\cite{bao02c}] to extrapolate to $E$=10 meV, 
where spin fluctuations were observed in La$_2$Cu$_{0.9}$Li$_{0.1}$O$_4$
using a thermal neutron spectrometer\cite{bao99a}.

$I^{3D}$ and $I^{q3D}$ appear simultaneously below $\sim$150~K.
Their concave shape in Fig.~\ref{fig4} differ drastically 
from the usual convex-shape of a squared order parameter, orange circles,
which was observed in $\mu$SR study below $T_g$=8~K\cite{Li214phs}. 
They are typical neutron scattering signal from  
slow {\em dynamic} spin correlations in spin-glasses\cite{FeAl_msm,sg_mh},
which fluctuate in the frequency window between 1 MHz 
and 17 GHz for $T> 8$~K, and below 1 MHz for $T< 8$~K.
Previously, energy-resolution-limited
neutron scattering from  
La$_{1.94}$Sr$_{0.06}$CuO$_4$ was observed to
have a similar temperature dependence as $I^{3D}$ in 
Fig.~\ref{fig4} and was attributed to
spin freezing\cite{bjbquasi}.
The kink of $I^{3D}$ at 20~K reflects an increased $T_g$ from 8~K 
to 20~K when probing frequency is increased from 1 MHz 
to 17 GHz\cite{bjbquasi,byoung}. At 0 Hz, $T_g\approx 6$~K
from DC magnetization measurements\cite{Li214phs2}. The increase
of $T_g$ with measurement frequency is a hallmark of glassy
systems\cite{byoung}.

\section{Discussions and summary}

The fact that $I^{3D}$ decreases below $T_g$(17GHz)$\approx$20~K 
while $I^{q3D}$ continues to increase indicates that 
the ``Edwards-Anderson order parameter''\cite{byoung,FeAl_msm,sg_mh} 
distributes only 
along lines such as the (1$k$0). In conventional spin-glasses, 
the ``Edwards-Anderson order parameter'' is more
isotropically distributed in the {\bf q}-space\cite{byoung,sg_mh,FeAl_msm}.
Thus, the spin-glass state in La$_2$Cu$_{0.94}$Li$_{0.06}$O$_4$
is characterized mainly by interlayer disorder which upsets phase correlation
between large antiferromagnetic clusters in different CuO$_2$ planes. 
This picture offers a possible alternative to the conventional
competing antiferromagnetic/ferromagnetic interaction model for
spin freezing in doped cuprates. In addition, it suggests that
the weak interlayer exchange interaction likely plays an important 
role in the finite temperature spin-glass transition in the quasi-2D 
Heisenberg magnetic systems.

Another important difference from conventional spin-glasses in which 
all spins are believe to freeze at low temperature is that only a fraction of spins
freeze in La$_2$Cu$_{0.94}$Li$_{0.06}$O$_4$. 
Other spins in 2D correlations 
remain fluctuating down to 1.5~K. This is consistent with 
numerical evidence that quantum fluctuations prevent spin-glass
transition for 2D $S$=1/2 Heisenberg system\cite{bhatt}.
The spin-glass component in our sample has to acquire interlayer
correlations to achieve a higher dimension in order to be realized.
It appears that the lower critical dimension for a $S$=1/2 Heisenberg
quantum spin glass is between 2 and 3.

A further difference from conventional spin-glasses, for 
which one can measure the narrowing of magnetic 
spectrum toward $E$=0\cite{FeAl_msm}, is that
when $S^{3D}({\bf q},E)$ and $S^{q3D}({\bf q},E)$ in 
La$_2$Cu$_{0.94}$Li$_{0.06}$O$_4$ become detectable at about 150~K,
they are already energy-resolution-limited, with spins fluctuating
much slower than 17 GHz.
This property of $S^{3D}({\bf q},E)$ and $S^{q3D}({\bf q},E)$ resembles
the classic central peak phenomenon in the soft phonon
transition\cite{SrTiO,SrTiOr}.
The disparate dynamics of the central peak and phonon are explained by
Halperin and Varma\cite{hcav} using a phase separation model:
defect cells contribute to the slow relaxing central peak
while coherent lattice motions (phonons) to the resolved 
inelastic channel. 
This mechanism has been applied with success to a wide class of 
disordered relaxor ferroelectrics\cite{Courtens82,Burns83}.

For La$_2$Cu$_{0.94}$Li$_{0.06}$O$_4$,
we envision that disorder accompanying doping
prevents the long-range order of the antiferromagnetic phase mainly by
upsetting interlayer magnetic phase coherence,
see Fig.~\ref{fig3} for the {\bf q}-distribution of
frozen spins. This upsetting is not uniform in the Griffiths
fashion\cite{qsg_hy} with weak and strong coupling parts in the 
sample. In our laminar material, however, the weak and strong coupling
parts have different dimensionality: 2D and nearly 3D, respectively.
The 2D part is a spin liquid and represents essentially the whole 
system at high temperature, see Fig.~\ref{fig4}. 
Part of sample with stronger interplane coupling tends to order 
three dimensionally below $\sim$150~K, producing
$S^{3D}({\bf q},E)$ and $S^{q3D}({\bf q},E)$. 
The condensation of the 2D spin liquid at $\sim$ 150~K 
into the quasi-3D dynamic clusters
of diminishing energy scale, instead of a true long-range order,
may reflect the divergent fluctuations which destabilize
static order at finite temperature
for 2D random $XY$ or Heisenberg systems\cite{bhatt,Hertz79,2d_gls}.
The nearly 3D spin-glass instead of a 3D antiferromagnet finally orders
at a much reduced $T_g\approx 20$~K, when $I^{q3D}+I^{3D}$
approaches the 2D spectral weight (Fig.~\ref{fig4}).
The coexistence of spin liquid and spin glass
components at low temperatures may be a general 
consequence of no ``mobility edge'' 
separating finite and infinite range correlations for a 2D
random system\cite{Hertz79}.
Recently, Monte-Carlo simulations of a doped 2D classical
antiferromagnet suggest that there are two populations of 
spins: one with fast and the other with slow dynamics\cite{sg_mpk}.
This is consistent with our experimental results and the 
Griffiths picture for random magnetic systems. A
phenomenological Halperin and Varma model may be built
for spin dynamics in doped cuprates based on these microscopic
insights.

In summary, spins in La$_2$Cu$_{0.94}$Li$_{0.06}$O$_4$
develop {\it dynamic} antiferromagnetic order in the CuO$_2$ plane 
with very long $\xi_{\Box}$ below 180~K. The characteristic
energy of the 2D spin fluctuations is 0.18$k_B T$ for $T>50$~K and
1~meV for $T<50$~K\cite{bao02c}. Below $\sim$150~K,
interlayer phase coherence appears between some of these planar
antiferromagnetic clusters with an energy scale smaller than 70~$\mu$\,eV.
While the 2D antiferromagnetic correlations in an individual plane remain
liquid down to 1.5~K, coherent multiplane
antiferromagnetic correlations become frozen below $T_g$.
The phase separation into 2D spin-liquid and spin-glass
of higher dimension with an unusual {\bf q}-structure for 
the ``Edwards-Anderson order parameter'' is most likely 
related to quasi-2D nature of magnetic exchange in 
the cuprates and is distinctly different from conventional spin-glasses. 

A theory of spin-glass in doped cuprates should include 
interlayer coupling. Theory explaining both the partial spin freezing and
the observed crossover\cite{bao02c,bao04a} of quantum
spin fluctuations are called for. The heterogeneous magnetic 
correlations, instead of a uniform magnetic phase,
suggests the possibility that superconductivity and the almost
3D antiferromagnetic order may reside in 
different phases in La$_{2-x}$Sr$_x$CuO$_4$ 
and Y$_{1-x}$Ca$_x$Ba$_2$Cu$_3$O$_{6+y}$.
Similar, detailed {\bf q}, $E$ and $T$ dependent cold neutron
spectroscopic study on these cuprates are desirable.

We thank R.H.\ Heffner, P.C.\ Hammel, S.M.\ Shapiro, C. Broholm,
L.\ Yu, Z.Y. Weng, A.C. Castro Neto, O. Sushkov, J. Ye,
F.C. Zhang, X. G. Wen, T. Senthil, P. C. Dai and C.M. Varma for useful discussions. 
SPINS and NG7-SANS at NIST are supported partially by NSF. Work at LANL is
supported by U.S. DOE.

\end{document}